# Tilted Micro Air Jet for Flow Control

Julien Malapert*, Réda Yahiaoui°, Rabah Zeggari and Jean-François Manceau
Institut Femto-st, Université de Franche-Comté, CNRS UMR 6174, département MN2S
32 Avenue de l'Observatoire
25044 Besançon Cedex
FRANCE
*julien.malapert@femto-st.fr ; °reda.yahiaoui@femto-st.fr

**Abstract:** In this paper, we present an interesting method to microfabricate a tilted micro air jet generator. We used the well-know deep reactive ion etching (DRIE) technique in order to realize in a silicon substrate a double side etching. For aircraft and cars, micro air jets will take an important place for fluid control. Micro air jets are characterized by their speed, frequency and tilt. Usually, this micro air jets are produced by fluidic microsystems. We presented experimental results about micro tilted air jets. A comparison between finite element method simulation, theory and experimental results are performed to define the microsystem geometry leading a specific air jet angle.

## 1. Introduction

In aeronautic field, one tries to control airflow close to aircraft wing [1]. A new method consists in producing micro air jets coming from wing surface to avoid turbulences and in order to improve aircraft aerodynamic performances [2]. These micro air jets have several characteristics such as speed, frequency and tilt. We fabricated on a silicon substrate a passive fluidic microsystem able to produce a tilted micro air jet [3]. The specific air jet angle depends on channel structure and geometry of the micro channel obtained by the well-know deep reactive ion etching (DRIE) microtechnique [4]. Instead of etching a tilted channel we made a double side deep reactive ion etching providing a tilted aperture which goes to deflect air flow like in a tilted channel case. For airflow control in aeronautic fields, one try to obtain an air jet making a 45° angle from the aircraft wing surface.

In this paper, we made comparison between finite element method simulation, theory and experimental results to define the microsystem geometry leading a specific air jet angle.

## 2. Theory

### 2.1 Air jet speed

According to rectangular section channel dimensions (several tens by hundreds of microns), microfluidics theory, taking into account the effects of depletion and the effects of unsteady gas, does not apply. The classical fluid dynamics theory governs the behavior of fluid inside the channels of our fluidics system. However, the equations of Navier and Stokes, involving the effects governing the incompressible fluid dynamic does not seem to be the easiest solution to estimate the flow speed inside the channels of our microsystem. The difference of pressure established between the entrance port and the ejection port allows quickly and coarsely to reach out the average speed of air jet at the ejection port ($v_{out}$) according to the formula stemming from the experiment of the tube of Pitot. This formula gives the average fluid speed according to the dynamic pressure at the tube entrance and the static pressure outside of the tube:

$$v_{out} = \sqrt{\frac{2}{\rho_{air}} \Delta P}$$

with $\rho_{air}$ = 1.2 Kg/m$^3$ at 20°C and $\Delta P = |P_{in} - P_{out}|$

### 2.2 Tilted air jet generator

Unlike conventional machining techniques, the collective silicon microfabrication of tilted hole is considered as a difficult task. Several micro techniques (wet or dry etching) like laser machining [5], FIB (Focused Ion Beam) machining [6], erosion by electrical discharge [7] or plasma etching (DRIE) are often used to obtain holes in silicon substrate. It is rather difficult to make tilted hole with these kinds of micro machining techniques because etching is essentially normal to the surface or they are not collective. Even if we can believe a wet isotropic silicon etching (KOH, TMAH) could be used to create inclined hole collectively, it unfortunately can not overcome the natural etching directions imposed by the crystalline mesh (typically 54.7° and 35.3° for a (100) silicon wafer). This fact limits the possibilities for hole angle.

A pressure difference established between every extremity of a channel allows the creation of an air flow going through the channel. The slope of the channel allows obtaining a tilted air jet. In our case the oblique jet is not obtained through a tilted channel but through a tilted window. The particularity of this opening allows tilting the jet as in the case of a tilted channel. The supposed jet slope depends on the both $d_1$ and $d_2$ etching depths as shown in figure 1. The theoretical air jet angle $\theta$ is proportional to the $z$ covering depth of front and back side etching, the $x$ opening and the thickness $T_w$ of the silicon wafer. We notice that:

$$z = d_1 + d_2 - T_w$$

$$\theta = \arctan\left(\frac{x}{z}\right)$$

$$\theta = \arctan\left(\frac{x}{d_1 + d_2 - T_w}\right)$$

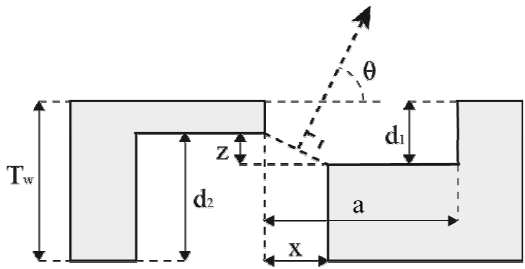

**Figure 1.** Sketch of the tilted micro air jet generator.

## 3. Numerical model

To study air jet characteristics, FEM (Finite Element Method) simulations on COMSOL software were conducted on a different air jet generator structures and for different boundary conditions.

### 3.1 Tilted air jet generator model

The model geometry consists of a horizontal flow channel in the middle of which is an aperture, a narrow diagonal structure (Figure 2). The fluid comes from left to top-right and goes through a tilted narrow aperture in the middle of the channel. This tilted window imposes a specific direction for air flow creating a tilted air jet at the channel exit.

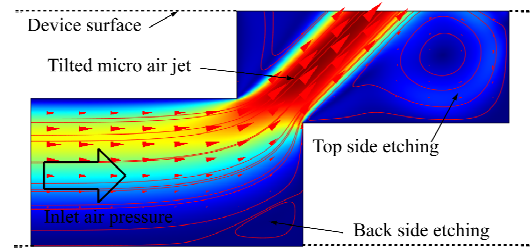

**Figure 2.** Plot of FEM results and tilted air jet generator structure.

### 3.2 Governing equation

The fluid flow in the channel is described by the Navier-Stokes equations, solving for the velocity field u = ($u$, $v$), the pressure p and the volume force F affecting the fluid. We assume no gravitation or other volume forces affecting the fluid, so F = 0. At the channel entrance on the left, the flow has fully developed laminar characteristics with a quasi-parabolic velocity profile because the condition p > 0 is applied. At the outflow (top boundary), the boundary condition is p = 0.

### 3.3 Numerical results

**Numerical results: air jet angle**

To study air jets tilt, several numerical calculations were conducted on air jet generator having different structure geometry. Each devices is composed of the same channel dimensions except tilted aperture dimensions ($x$ and $z$ dimensions). The simulation takes into account the volume of air in the cavities located on the top and the bottom of the device (figure 2). We can see easily the outgoing tilted air in the cavity located in the top side. We estimate that over this cavity, the jet retains its slope (50° in figure 2) to a distance several times equal to $d_1$ depth. While we neglect the viscosity of fluid for the estimation of theoretical jet speed, the simulation must take into account this parameter because it is this effects that will give the slope to the air jet which can be measured. Numerical calculation on different model having various x/z ratio was conducted. Measured angle are listed in the following table and where wee can see a quiet good agreement (~20%) between theory and numerical results.

| x/z | Theory angle (°) | Comsol angle (°) |
|-----|------------------|------------------|
| 2.4 | 67 | 53 |
| 1.4 | 54 | 47 |
| 0.83 | 40 | 32 |
| 0.52 | 28 | 24 |
| 0.40 | 23 | 23 |

**Numerical results: air jet speed**

In order to estimate the real air jet velocity, we wanted to compare numerical and theory results. We made numerical calculations with a specific structure where we applied a range of inlet pressure. We measured velocity at the exhaust port of the channel for several inlet pressures.

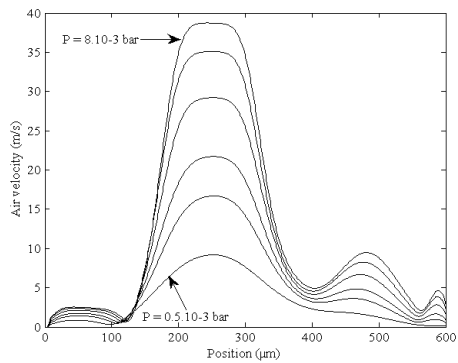

**Figure 3.** Air velocity along the exhaust port of the channel for different inlet pressure.

Maximum velocity is obtained at x = 250 µm for each inlet pressure. It seems the jet slope does not depend on the pressure. From these numerical results, it is possible to compare theory and numerical results. We can observe in figure 4 the good agreement between the both results and we can estimate real air jet velocity is to close them.

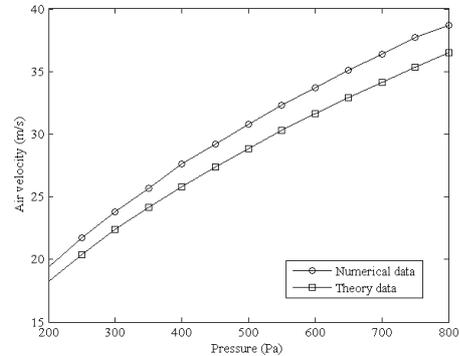

**Figure 4.** Air jet velocity (numerical and theory data) at the exhaust port of the channel for different inlet pressures.

## 4 Experiments

### 4.1 Experimental set up

In order to visualize the airflow, we use smoke to make air visible for video camera. The smoke comes from a smoke generator (figure 5). Smoke makes appear tilted micro air jet above jet generator.

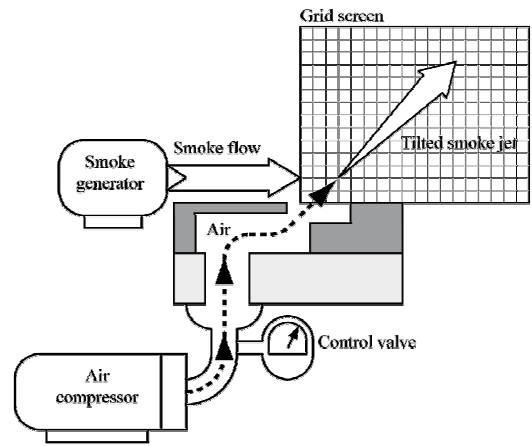

**Figure 5.** General structure of test bench for tilted air jet generators.

Air jet is obtained applying a minimal pressure of 0.5 bar until a maximum pressure of 2 bar in order to avoid air jet generator destruction. No variations of jet angle have been seen in function of inlet pressure variation like in numerical results. We can consider that air jet angle does not depend on air inlet pressure in our case.

**4.2 Experimental results**

Fluidic experiments are made at a pressure of 1 bar. Jet angles can be measured following two methods. First one, we can use a grid behind air jet (figure 6) and second one, we can use software able to measure angle from acquired pictures.

| x/z | Theory angle (°) | Measured angle (°) |
|------|------------------|--------------------|
| 2.4  | 67               | 53                 |
| 1.4  | 54               | 47                 |
| 0.83 | 40               | 32                 |
| 0.52 | 28               | 24                 |
| 0.40 | 23               | 23                 |

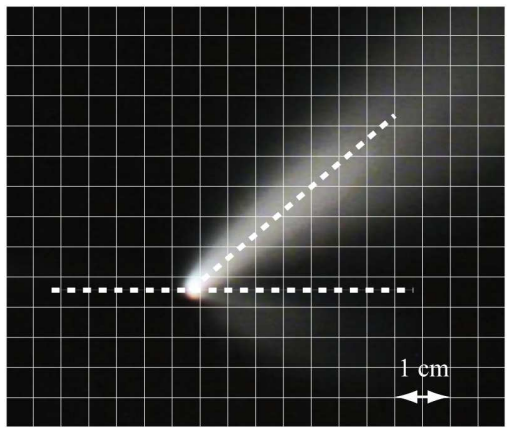

**Figure 6.** Picture of tilted micro air jet above jet generator (42°).

In the present case, according to generator architecture dimensions, obtaining a jet angle of less than 23° is impossible as well as an angle greater than 90°. Overall, between these two extremes, for theory, simulation and experiment, the more x/z ratio becomes high the more jet angle increases and vice versa (figure 7). We note that theory data seems to be a limit which can not exceed FEM simulation and experimental results. For example and at first glance, when x/z = 1, corresponding to a theoretical angle of 45°, this angle is approximately obtained by simulation only for x/z = 1.3 and for x/z = 1.6 in the case of experimental results.

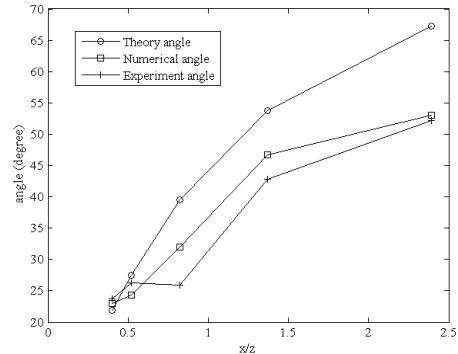

**Figure 7.** Air jet angle as a function of x/z ratio for theory, numerical and experimental results.

## 5. Conclusions

FEM calculation results combined with some fluidic experiments and applying a correcting factor allow defining structure geometry to generate a specific air jet angle. A 45° jet angle is not obtained with an x/z ratio equal to 1 but with a smaller ratio (1.3 and 1.6 for numerical and experimental results). The good agreement between theory and FEM simulation results involving air jet velocity allow us to estimate real air jet velocities for greater pressures. In the future, specific measurement technique of the velocity will be implemented to confirm calculations. Moreover, we will try to make systems using this kind of air jet generators to prove their efficiency for flow control. We want to precise we employed an original and easy method using the well-known DRIE technique to develope and to collectively make fluidic microsystems able to produce tilted micro air jets.

## 7. Acknowledgements

The authors would like to thank the technology staff of MIMENTO micro technology center, in Besançon, France and this research is funded by ANR Project 06 ROBO 0009 03.